\documentclass[12pt,showpacs,amssymb,superscriptaddress]{revtex4}
\usepackage{graphicx}
\usepackage{amsmath,amsthm,amsfonts}

\begin{document}
\title{Loop Quantum Cosmology: A cosmological theory with a view}
\author{G A Mena Marug\'an}
\email{mena@iem.cfmac.csic.es}
\address{Instituto de Estructura de la Materia, CSIC,
Serrano 121, 28006 Madrid, Spain.}

\begin{abstract}
Loop Quantum Gravity is a background independent, nonperturbative approach
to the quantization of General Relativity. Its application to models of
interest in cosmology and astrophysics, known as Loop Quantum Cosmology,
has led to new and exciting views of the gravitational phenomena that took
place in the early universe, or that occur in spacetime regions where
Einstein's theory predicts singularities. We provide a brief introduction
to the bases of Loop Quantum Cosmology and summarize the most important
results obtained in homogeneous scenarios. These results include
a mechanism to
avoid the cosmological Big Bang singularity and replace it with a Big Bounce,
as well as the existence of processes which favor inflation.
We also discuss the extension of the frame of Loop Quantum Cosmology to
inhomogeneous settings.
\end{abstract}
\pacs{04.62.+v, 04.60.-m, 98.80.Qc}
\maketitle

\section{Introduction}

Gravity remains as the only fundamental physical interaction
which has not been satisfactorily described quantum
mechanically. Leaving aside the belief that
all interactions should be unified
in a single theory, a strong motivation for
a quantum theory of gravity comes from the singularity
results of General Relativity \cite{hawk}. In these
classical singularities the predictability
breaks down, indicating that the regime of applicability of Einstein's
theory has been surpassed and that a new and
more fundamental description is needed.

Apart from the ability to cure the classical singularities,
one expects that a quantum theory of gravity would open a new window
to physics, incorporating phenomena which originate in the quantum realm.
Any candidate theory should make this compatible with
the infrared behavior of General Relativity, so that no quantum effect
alters much spacetime regions like those which we observe \cite{ash}.
Explaining why our Universe is actually so (semi-)classical
constitutes a real challenge to quantum gravity. Previous attempts to construct
a quantum formalism for the metric fields (in geometrodynamics) employing a
Wheeler-De Witt approach have proven unsuccessful. In the general theory,
the approach finds functional analysis and interpretational obstacles which
prevent further progress. On the other hand, in simple models where these
obstacles can be circumvented, the classical singularities are not fully
resolved (e.g., semiclassical states are peaked on trajectories where
physical observables eventually diverge \cite{aps,aps2}). Recently, a different
approach to quantize General Relativity has been developed: the theory of
Loop Quantum Gravity (LQG) \cite{al,thie}. This theory is based on a
nonperturbative canonical formalism of gravity and declares diffeomorphism
invariance and background independence to be basic guidelines.
Its application to simple cosmological models, generally homogeneous ones,
gives rise to the new area of gravitational physics known as Loop Quantum
Cosmology (LQC) \cite{ash,bojo}.

\section{Loop Quantum Gravity}

LQG is a canonical quantization of General Relativity, constructed starting from
a Hamiltonian formulation of Einstein's theory. Let us
begin by considering a globally hyperbolic spacetime. Introducing then a global foliation,
the initial data for the construction of the Lorentzian
spacetime metric are contained in the spatial 3-metric and the extrinsic curvature
on the considered section of the foliation \cite{wald}. If one wants to introduce fermions
in this framework, the spatial metric must be replaced with a triad, to which fermions
couple directly. This coupling occurs in an internal su(2) index, and respects
the invariance of the system under
internal rotations, realized then as SU(2) gauge transformations.
The spatial metric is recovered as the square of the triad, contracted
in the internal indices by means of the Euclidean metric
[the Killing-Cartan metric on su(2)].
A canonical set of variables to describe the gravitational
phase space is formed, up to numerical factors, by the densitized triad (i.e.,
the triad multiplied by the square root of the determinant of the spatial metric),
and the extrinsic curvature expressed in triadic form (i.e., the extrinsic curvature
contracted with the triad in one spatial index).

At this stage, one can replace the extrinsic curvature by a connection valued
1-form, taking values on the algebra su(2).
Classically, this Ashtekar-Barbero connection is a sum of the spin connection
compatible with the co-triad and the triadic extrinsic curvature
multiplied by an arbitrary positive number $\gamma$, known as the Immirzi
parameter. By construction, this connection forms  a canonical set
with the densitized triad, modulo a factor $8\pi G\gamma$ in their
Poisson brackets, where $G$ is the Newton constant (from now on, we set
the speed of light equal to the unity).

The introduction of the su(2)-connection
allows one to incorporate nonperturbative techniques in the description of the system,
similar to those developed in gauge field theories, like e.g. for Yang-Mills fields.
In particular, the gauge invariant information about the connection is captured in
the so-called Wilson loops, constructed from holonomies which describe the parallel
transport of
spinors around loops. We hence replace the connection by SU(2)-holonomies along
(piecewise analytic) edges, where we understand that an edge is an embedding of the
closed unit interval in the considered spatial manifold. This replacement involves
a 1-dimensional smearing of the connections, and
renders the information about them gauge invariant except for the effect of
transformations at the end points of the edges. By joining a finite number of
edges in those vertices to form a graph \cite{al}, and combining the holonomies
there so that SU(2) invariance is respected everywhere, one obtains what is
usually called a spin network. It is worth noticing that the construction of
the holonomies, which contain a line integral of the connections, is made
without recurring to any background structure.

Since the most relevant field divergences in our formalism come from the
appearance of a 3-dimensional delta function on the Poisson brackets between
the connection and the densitized triad, and we have already smeared the connection
over one dimension, it seems natural to smear the triad similarly, but now
over two dimensions. Given that the densitized triad is a (spatial) vector density,
this smearing can be carried out again without employing any background structure.
For any (piecewise analytic) surface, we can define the flux of the densitized
triad through it, obtaining the desired smearing. Holonomies and fluxes form an
algebra under Poisson brackets, which we regard from now on as our basic algebra
of functions on the gravitational phase space. From this perspective, the
quantization of General Relativity amounts to the representation of this algebra
as an algebra of operators acting on a Hilbert space. In addition, we must take
into account that the system is subject to a series of constraints, which must be
imposed quantum mechanically. These are  the Gauss [or SU(2)] constraint, the
diffeomorphisms constraint, and the Hamiltonian or scalar constraint, which express
the invariance of the system under SU(2) transformations, spatial diffeomorphisms,
and time reparametrizations \cite{al,thie}.

An important result for the robustness of the predictions of LQG is a uniqueness
theorem about the admissible representations of the holonomy-flux algebra, known as
the LOST theorem (after the initials of its authors \cite{lost}).
Specifically, this theorem states that there exists a unique cyclic representation
of that algebra with a diffeomorphism invariant state (interpretable as a "vacuum").
In total, the choice of the algebra of elementary variables, motivated by background
independence, and the status of spatial diffeomorphisms as a fundamental symmetry pick
up a unique quantization, up to unitary equivalence and prior to the imposition of
constraints.

To gain insight into the kind of quantization adopted in LQG, let us consider the
so-called cylindrical functions:
complex functions that depend on the connection only via the
holonomies along a graph, formed by a finite numbers of edges.
Completing this algebra of functions with respect to the norm of the
supremum, we obtain a commutative $C^*$-algebra with identity, where the $*$-relation
is provided by the complex conjugation \cite{curdi,vel}. According to Gel'fand theory,
this algebra is then (isomorphic to) the algebra of continuous functions on a
certain compact space, called the spectrum. Smooth connections are dense in this space.
Besides, the Hilbert space of the representation is just a space of square
integrable functions on this Gel'fand spectrum with respect to a certain measure.
The LOST theorem guaranties that there exists only one diffeomorphism invariant measure
which supports not only a representation of the holonomies, but of the whole holonomy-flux
algebra: the Ashtekar-Lewandowski measure, used to construct the representation in LQG.
This representation turns out not to be equivalent to a standard one, and therefore leads
to physical results which are different from those of other, conventional quantizations.
In fact, the representation is not continuous; as a consequence, the connection cannot
be defined as an operator valued distribution \cite{al}. Finally, the resulting quantum
geometry is discrete, with area and volume operators that have a point spectrum \cite{al}.

\section{Loop Quantum Cosmology: the flat FRW model}

As a paradigmatic example in LQC, let us now apply this type of loop
quantization techniques to Friedmann-Robertson-Walker (FRW)
cosmologies (namely homogeneous and isotropic spacetimes) with flat spatial sections
of $\mathbb{R}^3$ topology and a matter content provided by a homogeneous massless
scalar field $\phi$, minimally coupled to the metric \cite{aps,aps2}.
We introduce a fiducial triad and an integration cell, adapted to it, to carry out
all integrations and avoid in this way divergences due to the homogeneity and noncompactness of
the spatial sections. We call $V_0$ the fiducial volume of this cell. It is possible to
check that all physical results are indeed independent of these choices of fiducial
elements \cite{aps,aps2}. Besides, one can fix the gauge freedom
so that both the densitized triad and the connection become diagonal. Given the
isotropy, they are then totally specified by one single variable each, which we call $p$ and $c$,
respectively. These variables describe the geometry degrees of freedom, vary only in time, and
form a canonical pair: $\{c , p\}={8\pi \gamma G}/{3}$.
Classically, they are related with the scale factor $a$ and its time derivative by the formulas
$p= V_0^{2/3} a^2$ and $c=\gamma V_0^{1/3} \dot{a}$.

To retain all the gauge invariant information about the su(2)-connection,
taking into account the homogeneity, it suffices to consider holonomies along (fiducial)
straight edges. Similarly, triads are now smeared across (fiducial) squares.
The fluxes are then totally determined by the variable $p$. Returning
to the holonomies, it is easy to check that, for an edge of coordinate length
$\lambda V_0^{1/3}$ in any fiducial direction, the matrix elements of the SU(2)-holonomy
are linear combinations of exponentials of the form $e^{\pm i \lambda c/2}$.
The corresponding configuration algebra is then the
linear space of continuous and bounded complex functions in the real line ($c\in\mathbb{R}$),
with elements of the form $f(c)=\sum_j f_j e^{i\lambda_j c/2}$.
It is well known that the completion of this algebra with the supremum norm is just
the Bohr $C^*$-algebra of almost periodic functions \cite{vel}.
Its Gel'fand spectrum is the Bohr compactification of the real line, $\mathbb{R}_{\rm Bohr}$ .
This compactification can be seen as the set of group homomorphisms from
the group of real numbers (with the sum) to the multiplicative group $\mathbb{C}$
of complex numbers of unit norm. Indeed, for every real number $c$
we have a homomorphism  $x_c:  \mathbb{R}
\rightarrow \mathbb{C}$ of this kind, namely $x_c(\lambda)=e^{i\lambda c/2}$.
Moreover, it is possible to see that the real line is
actually dense in $\mathbb{R}_{\rm Bohr}$, using the fact that our initial
configuration algebra separates points $c \in \mathbb{R}$ \cite{curdi,vel}.

The operation $x \tilde x (\lambda)= x(\lambda) \tilde x
(\lambda)$ provides a commutative group structure in
$\mathbb{R}_{Bohr}$. Since the group $\mathbb{R}_{\rm Bohr}$ is compact,
it has a (unique) invariant Haar measure.
The functions on $\mathbb{R}_{\rm Bohr}$ consisting in the evaluation
at a real point $\mu$ form an orthonormal basis in the corresponding Hilbert space
of square integrable functions with the norm defined by that measure \cite{vel}.
We designate each element in this basis with a ket $|\mu \rangle$. This basis
allows us to pass from our configuration representation, in which holonomies act by
multiplication, to a ``momentum'' representation in which the triad has a multiplicative
action \cite{curdi,vel}. Calling
$N_{\lambda}=e^{i\lambda c/2}$,  this ``momentum'' representation is given by
$\hat{p}|\mu\rangle=({4\pi\gamma G}/{3})\mu|\mu\rangle$ and $\hat{N}_{\lambda}|\mu\rangle=
|\mu+\lambda\rangle$ (we set $\hbar=1$).
Clearly, the basis $\{|\mu \rangle;\;\mu\in\mathbb{R}\}$ is
uncountable, and therefore the Hilbert space is nonseparable. Nevertheless,
normalizable states can get nonvanishing contributions only from a
countable subset of states $|\mu \rangle$; otherwise their norm would not be finite.
This ``momentum'' representation
is the one usually employed in LQC.
It is worth remarking that the representation fails to be continuous, owing to the discrete
norm on the Hilbert space, $\langle \mu | \tilde \mu \rangle = \delta^{\tilde \mu}_{\mu}$.
As a result, a connection operator does not truly exist, and
the representation is inequivalent to the Wheeler-De Witt one, in total parallelism
with the situation found in LQG for the general case.

Although homogeneity ensures that the diffeomorphisms constraint is satisfied, and the
$SU(2)$ constraint has been removed by gauge fixing, the system is still subject to a Hamiltonian
constraint, which must be imposed now quantum mechanically. In order to introduce a Hamiltonian constraint
operator, there are essentially two building blocks which must be defined in terms of our elementary
operators $\hat{p}$ and $\hat{N}_{\lambda}$. First, we need an operator to
represent the phase space function $t(p)={\rm sign}(p)/\sqrt{|p|}$.
This function contains all the $p$-dependence of
the (nondensitized) triad of the model. Note that this triad diverges at the Big Bang, where the
variable $p$ vanishes. Correspondingly, the operator
representing $t(p)$ cannot be defined exclusively from $\hat{p}$ using the spectral theorem:
$\hat{p}$ has a point spectrum which contains the zero, and hence its inverse operator is not well defined.
But it is possible to construct a regularized triad operator using commutators with holonomies,
in addition to $\hat{p}$ \cite{abl}: $\widehat{t(p)}
=3(\hat{N}_{-\bar{\mu}} |\hat{p}|^{1/2} \hat{N}_{\bar{\mu}}-
\hat{N}_{\bar{\mu}} |\hat{p}|^{1/2} \hat{N}_{-\bar{\mu}})/(4\pi\gamma G \bar{\mu}).$ In principle, $\bar{\mu}$
may take any real value. It will be fixed later on in our discussion.
The resulting operator is diagonal in the considered basis of $\mu$-states.
Furthermore, it turns out to be bounded from above, so that, in particular, the classical divergence
disappears. Actually, our regularized operator annihilates the kernel of $\hat{p}$.

The other block that we need is the SU(2)-curvature operator. Recall that the connection operator is not
well defined, therefore we cannot use it to construct the curvature. Nonetheless, it is possible to
determine it using a square loop of holonomies. We use again edges of fiducial length $\bar{\mu} V_0^{1/3}$.
Classically, the expression of the curvature would
be recovered exactly in the limit of zero area, when $\bar{\mu}$ tends to zero. However, this limit
is not well defined in LQC. The idea is to shrunk the square up to the minimum physical area $\Delta$
allowed in LQG, where the spectrum of the area operator is discrete \cite{aps2}.
This introduces a certain nonlocality in
the formalism, and turns the parameter $\bar{\mu}$ into a state dependent quantity, since the physical
area depends on the particular state under consideration. Explicitly, the relation that must be satisfied
for each state is $\bar{\mu}^2 |p| =\Delta$. At this stage, it is convenient to relabel the basis
of $\mu$-states as a basis of volume eigenstates, introducing an affine parameter $v$ for the translation
generated by $\bar{\mu} c/2$ \cite{aps2}. By construction, we then get that $\hat{N}_{\bar{\mu}}|v\rangle=
|v + 1\rangle$. The parameter $v$ is related with the physical volume of the fiducial cell, $V=p^{3/2}$,
by the formula $v= {\rm sign}(p) V/(2\pi \gamma G \sqrt{\Delta})$.

Finally, for the quantization of the matter field $\phi$, we use the standard Schr\"odinger representation.
So, the kinematical Hilbert space is the tensor product of the gravitational space of LQC and the
standard one for matter. With a suitable factor ordering and choice of densitization \cite{mmo},
one then gets a Hamiltonian constraint of the form $\hat{H}=-6 \hat{\Omega}^2+ \hat{P}_{\phi}^2$, where
$\hat{P}_{\phi}$ is the momentum operator of the matter field, and acts by differentiation
(namely, $\hat{P}=-i\partial_{\phi}$). The gravitational part of the constraint is given by the operator
$\hat{\Omega}^2$.
Remarkably, this constraint leaves invariant the zero-volume
state $|v=0\rangle$, as well as its orthogonal complement. Therefore, the analog of the
classical singularity is removed in practice from the Hilbert space, and we can restrict all
 physical considerations to its complement. In this sense, the singularity gets resolved at a kinematical level.
Moreover, the operator $\hat{\Omega}^2$ has an action of the following type:
$\hat{\Omega}^2 | v \rangle = f_+(v) |v+4 \rangle
+ f(v)  | v \rangle + f_-(v)  | v -4 \rangle $. Here, the real functions $f_+(v)$ and $f_-(v)$
have the outstanding property that they
vanish in the respective intervals [-4,0] and [0,4] \cite{mmo}.
Thus, the action of $\hat{\Omega}^2$ preserves each of the subspaces of the gravitational
Hilbert space obtained by restricting the label of the $v$-states to any of the
semilattices $\mathcal L_{\varepsilon}^\pm=\{v=\pm(\varepsilon+4n);
\,n\in\mathbb{N}\}$, where $\varepsilon\in(0,4]$.
Then, each of these semilattices provides a superselection sector. In each sector, the orientation
of the triad is definite ($v$ does not change sign) and $|v|$ has an strictly
positive minimum, equal to $\varepsilon$. For concreteness, we choose sectors with $v>0$ from now on.

On the other hand, it is possible to show that, on each sector, the operator
$\hat{\Omega}^2$ has a nondegenerate absolutely continuous spectrum equal to the positive real line \cite{mmo,mmp}.
Recalling its action, this gravitational constraint operator might be understood as a second-order difference operator.
But its eigenfunctions are entirely determined by their value at $\varepsilon$, point from which they can be constructed
by solving the eigenvalue equation. In this sense, the gravitational constraint operator leads to a No-Boundary
description, in which the eigenstates which encode the information about the quantum geometry are all determined
without the need to introduce any boundary condition in the region around the origin.
We also notice that, up to a global phase, these eigenfunctions $e^{\varepsilon}_{\delta}(v)$ are real, since so is
the gravitational constraint operator.

With such eigenfunctions, one easily finds the solutions to the Hamiltonian
constraint, which have the form $\psi(
v,\phi)=\int_0^{\infty} d\delta\, e^{\varepsilon}_{\delta}(v) [\psi_+(\delta) e^{i\sqrt{6\delta}\phi}
+ \psi_-(\delta) e^{-i\sqrt{6\delta}\phi}].$ The scalar field $\phi$ plays the role of an emergent
time. Then, physical states can be identified
(e.g.) with the positive frequency solutions $\psi_+(\delta)$ that are square integrable
over the spectral parameter $\delta\in\mathbb{R}^+$ \cite{mmo}. A complete set of Dirac observables is formed
by $\hat{P}_{\phi}$ and $|\hat{v}|_{\phi_0}$, the latter
being defined by the action of the
volume operator when the field equals $\phi_0$. On the Hilbert space of
physical states specified above, these observables
are self-adjoint operators.

\section{The Big Bounce}

In the previous section we have completed the quantization of the flat FRW with a homogeneous massless
scalar field. In order to analyze the physical predictions of this quantum theory, we will consider now
the evolution of (positive frequency) quantum states with a semiclassical behavior. We study {\sl Gaussian-like}
states which, at an instant $\phi=\phi_0$ in the region of large emergent times $\phi_0\gg 1$, are
peaked on certain values of the elementary observables of the model, namely, the matter
momentum, $P_{\phi}=P_{\phi}^0$, and the physical volume, $v=v^0$.
We restrict our attention to states with large values of $P_{\phi}^0$ and $v^0$ \cite{aps,aps2}.
The numerical analysis of the quantum evolution unveils an outstanding phenomenon in these states:
the Big Bang singularity is resolved dynamically and is replaced by a bounce that connects the universe with
another branch of the evolution, dictated again by the equations of General Relativity.
This mechanism to elude the cosmological singularity is known as the Big Bounce \cite{aps,aps2}.

The numerical studies show that the considered semiclassical states remain peaked on a well defined
trajectory during the whole evolution. On these states, the Big Bounce does not occur in a
genuinely quantum region where one were to loose an effective notion of geometry and spacetime.
The trajectory deviates from the one predicted by General Relativity only when the matter energy
density $\rho$ becomes of the order of one percent of a critical density,
$\rho_{\rm crit}$. This scale for the onset of corrections is of the Planck order and universal: it is the same
for all the semiclassical states which suffer the bounce. Explicitly, the critical density is
$\rho_{\rm crit}= ({\sqrt{3}}/{32\pi^2 \gamma^3 G^2})\approx 0.41 \rho_{\rm Planck}$, where $\rho_{\rm Planck}$
is the Planck density. For densities close to the critical one, i.e., in the regime close to the bounce,
gravity behaves as a repulsive force owing to the effects of quantum geometry \cite{ash}.

In addition, the trajectory followed by the peak of these states matches an effective dynamics, which has
been deduced in detail (under certain assumptions on the family of states under consideration)
using techniques of geometric quantum mechanics \cite{taveras}. The agreement between the numerical simulations
and the predictions of this effective dynamics is remarkable. In particular, the effective dynamics predicts a
bounce precisely when the matter density reaches the critical value $\rho_{\rm crit}$.
Further support to the role played by this critical density comes from the study of the operator which represents
the matter density in the quantum theory. It is possible to prove that it has a bounded spectrum, the bound being
given again by $\rho_{\rm crit}$. Then, the overview picture is that the emergence of important quantum geometry effects
in this model is controlled by the value of the matter energy density. When this density approaches the Planck
scale, quantum geometry phenomena enter the scene, preventing that it keeps on increasing and consequently avoiding
the collapse into a cosmological singularity. It is worth pointing out that these quantum phenomena can be relevant
even in regions which one would not consider to belong to the deep Planck regime. For instance, the
volume $v$ at the bounce is proportional to the value of the matter field momentum, which is conserved in
the evolution. Hence, when the bounce occurs, the volume can be as large as desired.

The presence of a quantum bounce is actually generic in this quantum model,
with implications that exceed the
restriction to the discussed class of semiclassical states. We have already commented that the eigenfunctions
of the gravitational constraint operator are real (up to a global phase). Studying the Wheeler-De Witt limit
of this operator, one can prove that its eigenfunctions lead in fact
to positive frequency solutions with ingoing and outgoing components of equal amplitude in this limit
\cite{mmo} (see \cite{acs} for the case of a specific superselection sector).

Furthermore, the Big Bounce mechanism is not restricted just to the flat FRW model
with a massless scalar field, but is rather general. On the one hand,
assuming the validity of the effective dynamical equations for other matter contents (assumption that is supported
by the numerical analyses carried out so far), one can show that all strong singularities (\'a la
Kr\'olak) are resolved in flat FRW for any kind of matter \cite{singh}.
Only Type II and Type IV singularities may remain \cite{singh}, but these singularities can be considered physically
harmless, since geodesics can be extended beyond them (then, sufficiently strong in-falling detectors can survive
these singularities). On the other hand, similar conclusions about the occurrence of the Big Bounce have been reached in
other FRW models quantized in LQC. These include the flat model with negative cosmological constant \cite{benpaw}, the
closed model \cite{apsv},
the open model \cite{vand} (some problems of the treatment presented in that reference
can be solved with the techniques of \cite{awe2}), and
the flat model with positive cosmological constant (recently studied by Ashtekar and Paw{\l}owski), all of them with
a homogenous scalar matter field present as well.
For the flat FRW universe with negative cosmological constant and the closed FRW model, the classical evolution leads to
a Big Crunch (i.e., the universe recollapses into a cosmological singularity). In the quantum theory, this Big Crunch is
also resolved via a Big Bounce, like in the case of the Big Bang. In all cases, there exists an upper bound for the matter
energy density, which is given again exactly by $\rho_{\rm crit}$, and the infrared regime shows an outstanding agreement with
General Relativity.

In spite of some statements that have appeared in the literature of LQC \cite{singh},
the effective equations for flat FRW in the presence of generic matter do not necessarily lead
to an asymptotically de Sitter behavior if a vanishing or a divergent value of the
scale factor $a$ were to be approached in the evolution, without further assumptions.
The confusion comes from the consideration of
the identity $\ln{(\rho/\rho_0)}=\int^{a}_{a_0} [1+w(\tilde{a})] d\tilde{a}/\tilde{a}$,
deduced from the conservation equation for the matter energy density, and where $a_0$ is a reference value
for the scale factor, $\rho_0=\rho(a_0)$, and $w(a)$ is the ratio
between the pressure and the energy density of matter. In fact,
the convergence of the above integral
when $a\rightarrow \infty$ does not need that $w(a)$ tend to minus the unity \cite{jara}
(the value that would correspond to a de Sitter regime). Besides, even if the energy density is
required to be positive and bounded from above by the critical density,
one may have a vanishing limit for it. Then, the considered
integral would diverge in that limit, allowing for values of $w$ different from minus one
\cite{jara}. This situation might be found in the limit of null $a$.

Similar conclusions about the actual resolution of cosmological singularities have been
reached in other homogeneous scenarios which contain anisotropies.
LQC has been implemented successfully to completion in Bianchi models of
type I \cite{mmp,chiou,awe,gmm,mmwe}, type II \cite{awe2}, and type IX \cite{we}.
The analysis, complemented with numerical simulations in these models, confirms the Big Bounce
scenario, though now there may exist bounces in different scale factors,
since the spatial directions are not all equivalent in view of the anisotropy.
Together with the BKL conjecture \cite{bkl}, these results suggest a generic resolution of
spacelike singularities in LQC. Recall that the BKL conjecture says that spatial derivatives
can be neglected against time derivatives when one approaches a spatial singularity, so that
the dynamics at any point can be approximated locally by a homogeneous model
(i.e., a Bianchi model).

The anisotropic model which has received more attention is Bianchi I.
Recently, this model has been studied thoroughly with the prescription put forward by Ashtekar and
Wilson-Ewing \cite{awe} to determine the lengths of the edges used to define the holonomies that enter
the Hamiltonian constraint. Actually, it has been shown that this prescription
is characterized by the requirement that the action of the corresponding holonomies produce a constant shift in the
(absolute value of the) physical volume \cite{gmm}. In addition, one can see that the
initial value problem for the quantum evolution is well posed even in vacuo: an infinite though countable
set of initial data on the section of minimum physical volume suffices to fix the solution to the
constraint \cite{mmwe}. This allows one to identify the space of solutions and construct
from it the Hilbert space of physical states. Finally, while the superselection sectors for the
physical volume are the same as in flat FRW, the sectors for the anisotropies
turn out to have a rather different structure. They are still discrete, and different
triad orientations are not mixed, but these sectors are dense on the
(positive) real line, instead of being formed by points separated by a constant distance \cite{gmm}.

Further support to the Big Bounce scenario in anisotropic models, using the prescription of
\cite{awe} in the quantization, comes from the extrapolation of
the effective dynamical equations, assuming their validity for Bianchi I.
These effective equations guarantee that the directional Hubble rates, the expansion, and the
shear scalar (of comoving observers) are all bounded from above in the evolution \cite{cs},
preventing in this way the formation of dangerous singularities. Contrary to what one could naively expect
\cite{cs}, however, the bounded nature of these physical quantities cannot be extrapolated to the genuine
quantum theory. The reason is that a bounded function on phase space (namely, one of our physical quantities
in the effective theory) is not always represented by
a bounded operator. One can ensure that the corresponding operator is bounded only when it can
be defined in terms of a set of commuting elementary ones via the spectral theorem. But in generic situations
this is not the case. Similarly, an unbounded function on phase space may be represented by a bounded
operator. In addition, when superselection sectors enter the scene, the physically relevant spectra
of our elementary operators do not coincide with the range of their classical analogs, therefore introducing
limitations to the domain of applicability of the effective equations. For instance, the spectrum of the
(absolute value of the) physical volume is bounded from below by a positive number on each of our superselection
sectors. This invalidates the analysis of the limit of vanishing volume in the effective equations.

\section{Inflation}

The effects of quantum geometry in the {\sl early universe} are important not only to elude cosmological
singularities, but also to build up a satisfactory inflationary scenario. In standard cosmology,
one generally needs a fine tuning of the initial conditions or the inflationary parameters in order to
reach enough inflation to explain the observed universe, with at least 68 e-foldings. Recent results
in LQC indicate that, on the contrary, the quantum phenomena that accompany the Big Bounce render natural
an inflationary process with this number of e-foldings.

Let us consider a flat FRW model with an inflaton field with positive kinetic energy.
The effective equations for this type of models in LQC imply a series of interesting properties
\cite{ash,sloan}. Firstly, the Hubble parameter is bounded from above.
Besides, when the inflaton potential is nonnegative and bounded from below, the time derivatives
of the inflaton and of the Hubble parameter turn out to be bounded from above in absolute value.
In addition, for potentials which are unboundedly large when the inflaton field approaches plus/minus
infinity, there exists an upper/lower bound on the value of this inflaton.
On the other hand, and more remarkably, there always exists a phase of superinflation \cite{bojoinf,singhinf}
after the bounce, in which the Hubble parameter increases from zero up to its maximum value.
This phenomenon of superinflation is robust in LQC, and appears even in the absence
of potential. Nevertheless, one can show that the superinflation
epoch alone does not yield sufficient e-foldings in generic situations.

For concreteness, we will focus our discussion on the inflationary potential $m^2\phi^2$,
with a mass of the order of $10^{-6}$ in Planck units, which is the phenomenologically preferred value.
We can calculate the probability of getting more than 68 e-foldings,
starting with equiprobability for every unconstrained set of initial data. Taking the bounce as
the most natural instant to define initial data, and imposing the Hamiltonian constraint, we obtain on the space of
physical data a measure of the form
$\sqrt{1-F_{\rm Bounce}}\, d\phi_{\rm Bounce}\, d v_{\rm Bounce}$, where $\phi_{\rm Bounce}$ and $v_{\rm Bounce}$
are the values of the field and the physical volume on the bounce section, respectively. Besides,
$F_{\rm Bounce}$ is the fraction of the matter energy density which is due to the potential at that moment.
On the other hand, we recall that $|\phi_{\rm Bounce}|$ is bounded,
because the studied potential grows unboundedly. Then,
it is possible to see that, if $F_{\rm Bounce}>1.4 \cdot 10^{-11}$,
the superinflationary phase either provides the desired number of
e-foldings by its own or supplements them by funneling the solutions to initial conditions
such that there is a sufficiently long period of slow-roll inflation \cite{sloan}.
It is straightforward to compute that the relative probability for a value of $F_{\rm Bounce}$
in this range is greater than 0.99. These conclusions are not sensitive to reasonable changes, e.g.,
in the inflaton mass or the form of the potential. Therefore, the effective equations that arise in LQC
solve the fine tuning problems for inflation.

\section{Inhomogeneous models}

Inhomogeneous models have been considered lately in the framework of LQC in order to extend the applicability
of this quantization scheme beyond the simple FRW or Bianchi cosmological spacetimes and with the aim at obtaining
predictions that, eventually, might be contrasted with observations.
The quantum analysis of this type of models
has been carried out adopting a hybrid approach \cite{gmm,mmwe,hybrid1,hybrid2}, which combines
a loop quantization of the
subspace of homogeneous solutions of the system (or rather of their geometry) and
a Fock quantization of the matter fields and inhomogeneous gravitational waves contained in the model.
This hybrid approach is based on the assumption that there exists a hierarchy of quantum phenomena, so that the
most relevant effects of the loop quantum geometry are those that affect the sector of
homogeneous degrees of freedom.

For a series of cosmological models,
the ambiguity in the selection of a Fock quantization for the
inhomogeneities can be removed by appealing to some recent theorems which guarantee the uniqueness of the choice of both a
field description (among a reasonable set of possibilities) and a Fock representation for it
\cite{uni1,uni2,uni3,uni4}. The uniqueness is ensured if the quantization respects certain conditions on
the unitarity of the dynamics as well as the invariance of the vacuum under the spatial symmetries of the
field equations. As a consequence,
the predictions of the hybrid quantization are robust, in the sense that they are not affected by the typical
ambiguities which plague the quantization of systems with an infinite number of degrees of freedom.
Cosmological systems where these uniqueness results have been proven, and the hybrid quantization is at hand,
include the Gowdy spacetimes \cite{gow} (with different spatial topologies
and possibly containing scalar fields), and fields and perturbations around closed FRW models.

In the inhomogeneous cosmologies that have been studied in this way, the loop quantization of the homogeneous
sector suffices to resolve the cosmological singularities at the kinematical level, in a similar manner as it
happens for the flat FRW model. It is remarkable that, in spite of the field character of these systems,
the quantization can been carried out to completion.
Moreover, the Hilbert space of physical states that one attains with this hybrid approach is such that
one recovers the Fock description of the inhomogeneities of the system, providing support to this conventional
quantum treatment and proving its compatibility with the loop quantization.

In particular, the hybrid quantization process has been discussed in detail
in the (vacuum) Gowdy model with 3-torus topology \cite{gmm,hybrid2}, probably the simplest inhomogeneous cosmology.
In this case, the effective dynamics that one obtains extrapolating the results deduced in FRW scenarios has been
studied both numerically and (partially) analytically \cite{bmp} (although the prescription used to determine the lengths
of the edges for the holonomies is not that in \cite{awe}).
The analysis confirms the presence of the Big Bounce, which happens in all the three spatial directions
of this anisotropic model. The bounce occurs typically at values of the fluxes variables (i.e., the $p$-variables for each
of the three spatial directions) which are at least a 13 per cent of those found in the absence of inhomogeneities \cite{bmp}.
This proves
that the inhomogeneities do not have a drastic effect that could alter substantially the Big Bounce, e.g., driving the
bounce into the deep Planck regime. In addition, the numerical studies have considered
the change in amplitude of the inhomogeneous modes
(which describe linearly polarized gravitational waves) between the two asymptotic
regions of the effective trajectories, which correspond respectively to a contracting and an expanding universe.
One can take a statistical average, disregarding phases in the mode decomposition of the inhomogeneities.
Actually, it is possible to see that, in the sector of the space of solutions where the inhomogeneities
dominate the bounce dynamics, the change in the amplitudes is antisymmetric with respect to the phase.
Then, in average, the amplitudes are statistically preserved through the bounce.
On the other hand, in the sector where the vacuum dynamics is approximately valid around the
bounce, the change in the amplitudes is positive in average \cite{bmp}.
Although the scenario is not completely physical, since the considered inhomogeneities are not all
those allowed in the most general cosmological setting,
the found behavior may indicate
a LQC mechanism that removes low amplitudes through the bounce.

\section{Conclusion}

The increasing attention paid recently to the quantization of cosmological systems using loop techniques has
crystalized in the foundation of a new branch of gravitational physics. This new formalism of LQC
allows one a rigorous control on the mathematical and interpretational
aspects of quantum cosmology, providing significance and robustness to the predictions in an area where they
are rarely falsifiable in a direct way. Moreover, in doing so, LQC has opened new views to the quantum phenomena of the
{\sl early universe}. It leads to a new paradigm for cosmology in which the Big Bang singularity is resolved and replaced with
a Big Bounce. Remarkably, this Big Bounce respects the semiclassicality of the universe, connecting two branches whose
asymptotic behavior is well described by General Relativity. On the other hand, LQC renders inflation a natural process,
suppressing the need for a fine tuning in the initial conditions or parameters of the inflationary era.
In addition, LQC might supply a mechanism to remove low amplitudes from the (primordial) inhomogeneities in cosmology.
And it suggests new settings for the consideration of initial conditions on the inhomogeneities,
which would not be imposed anymore in the region around the classical singularity.
Further research is needed to explore the consequences in issues such as cosmological perturbations, primordial
fluctuations, or the power spectrum of anisotropies in the cosmic background. These are exciting topics that
can provide a suitable arena where the predictions of LQC might be compared with observations.

\acknowledgments

The author is grateful to the organizers of the ERE2010 for
the warm atmosphere created during the conference. He thanks J. Cortez,
G. Jannes, L. J. Garay, M. Mart\'{\i}n-Benito, D. Mart\'{\i}n-de Blas, J. Olmedo,
and specially D. Jaramillo and J. M. Velhinho for enlightening conversations. Owing to space
limitations, many authors that
have contributed to the progress of LQC have not been cited explicitly.
The list of names includes (but is not restricted to)
K. Banerjee, G. Date, W. Kaminski, G. Khanna, G. M. Hossain, J. Lewandowski, C. Rovelli,
S. Shankaranarayaran, L. Szulc, M. Varadarajan, F. Vidotto, and Yongge Ma.
This work was supported by the MICINN Project No. FIS2008-06078-C03-03
and the Consolider Program CPAN No. CSD2007-00042.

\end{document}